\documentclass[twoside]{article}
\usepackage{fleqn,espcrc2,epsfig}

\newcommand{\AmS}{{\protect\the\textfont2
  A\kern-.1667em\lower.5ex\hbox{M}\kern-.125emS}}

\title{Dynamical next-to-next-to-leading order parton distributions and 
the perturbative stability of $F_L(x,Q^2)$}

\author{Cristian Pisano \address{Vrije Universiteit, Department of Physics and 
Astronomy, \\ De Boelelaan 1081, NL-1081 HV Amsterdam, The Netherlands}}

\begin{document}

\begin{abstract}
It is shown that the previously noted extreme perturbative
instability of the longitudinal structure function $F_L(x,Q^2)$ 
in the small Bjorken-$x$ region, 
is a mere artefact of the commonly utilized  `standard' gluon
distributions.  In particular it is demonstrated that using the 
appropriate  -- dynamically generated -- parton distributions at NLO
and NNLO, $F_L(x,Q^2)$ turns out to be perturbatively rather
stable already for $Q^2  \geq {\cal{O}}\, (2-3$ GeV$^2)$.
\end{abstract}

\maketitle
\begin{table*}[th]
\setlength{\tabcolsep}{1.pc}
\newlength{\digitwidth} \settowidth{\digitwidth}{\rm 0}
\catcode`?=\active \def?{\kern\digitwidth}
\caption{Parameter values of the NNLO and NLO QCD fits
with the parameters of the input distributions referring to (1)
at a common input scale $Q_0^2=\mu^2=0.5$ GeV$^2$ optimal at both perturbative
 orders.}
\label{tab:effluents}
\centering
\begin{tabular*}{\textwidth}{@{}l@{\extracolsep{\fill}}rrrrrrrrr}
\hline
& \multicolumn{4}{c}{NNLO}  & 
\multicolumn{4}{c}{NLO} \\
\hline
& $u_v$ & $d_v$ & $\bar{q}$ & $g$ &
  $u_v$ & $d_v$ & $\bar{q}$ & $g$ \\

N &  0.6210  &   0.1911     & 0.4393    &  20.281   &
     0.5312   &   0.3055     & 0.4810    &  20.649  \\

a &    0.3326      &  0.8678     & 0.0741    &  0.9737    &
       0.3161      &  0.8688     & 0.0506    &  1.3942       \\ 

b &    2.7254     &  4.7864      &  12.624   &   6.5186   &
       2.8205     &  4.6906      &  14.580   &   11.884       \\

c &    -9.0590     & 65.356        & 2.2121  & ---    &
       -8.6815     & 44.828        & -2.2622 &  15.879    \\

d &  53.547      &   1.6215        & 7.7450    &  ---      &
     54.994      &  -5.3645        &  21.650   &  ---    \\

e &   -36.979    &    -41.117        & --- & --- &
      -40.088    &     -21.839       & --- & --- \\

$\chi^2/{\rm dof}$ & \multicolumn{4}{c}{1.037} & 
                     \multicolumn{4}{c}{1.073} \\
$\alpha_s(M_Z^2)$ & \multicolumn{4}{c}{0.112} & 
                      \multicolumn{4}{c}{0.113} \\

\hline
\end{tabular*}
\end{table*}

A sensitive test of the reliability of perturbative QCD is provided by
studying \cite{ref1,ref2,ref3,ref4,Gluck:2007sq} the perturbative 
stability of the 
longitudinal structure function $F_L(x,Q^2)$ in the very small 
Bjorken-$x$ region, 
$x$ \raisebox{-0.1cm}{$\stackrel{<}{\sim}$} $10^{-3}$, 
at the perturbatively relevant low values of 
$Q^2$ \raisebox{-0.1cm}{$\stackrel{>}{\sim}$} ${\cal{O}}(2-3$ GeV$^2$).
For the perturbative--order independent rather flat toy model parton
distributions in \cite{ref1}, assumed to be relevant at $Q^2\simeq 2$ 
GeV$^2$, it was shown that next--to--next--to--leading order (NNLO)
effects are quite dramatic at 
$x$  \raisebox{-0.1cm}{$\stackrel{<}{\sim}$} $10^{-3}$
(cf.\ Fig.\ 4 of \cite{ref1}).  To some extent such an enhancement
is related to the fact, as will be discussed in more detail below, 
that the third--order $\alpha_s^3$ contributions to the longitudinal
coefficient functions behave like $xc_L^{(3)}\sim -\ln x$ at small $x$, as
compared to the small and constant coefficient functions at LO and NLO,
respectively.  It was furthermore pointed out, however, that at higher
values of $Q^2$, say $Q^2 \simeq 30$ GeV$^2$, where the parton
distributions are expected to be steeper in the small--$x$ region
(cf.\ eq.~(13) of \cite{ref1}), the NNLO effects are reduced
considerably.  It is well known that dynamically generated parton
distributions \cite{ref5} are quite steep in the very small--$x$ region
already at rather low $Q^2$, and in fact steeper \cite{ref6} than their
common  `standard' non--dynamical counterparts.  Within this latter
standard approach, a full NLO (2--loop) and NNLO (3--loop) analysis
morevover confirmed \cite{ref2,ref3} the perturbative fixed--order
instability expectations of \cite{ref1} in the low $Q^2$ region.

It is therefore interesting to study this issue concerning the perturbative
stability of $F_L(x,Q^2)$ in the low $Q^2$ region, 
$Q^2$ \raisebox{-0.1cm}{$\stackrel{<}{\sim}$} 5 GeV$^2$, within the
framework of the dynamical parton model \cite{ref5,ref6}.  For this 
purpose, following \cite{Gluck:2007sq}, 
 we repeat our previous \cite{ref8}  `standard' evaluation of
the NLO and NNLO distributions within the dynamical approach where the
parton distributions at $Q > 1$ GeV are QCD radiatively generated from
{\em{valence}}--like (positive) input distributions at an optimally
determined $Q_0\equiv \mu < 1$ GeV (where  `valence--like' refers to
$a_f >0$ for {\em{all}} input distributions $xf(x,\mu^2)\sim
x^{a_f}(1-x)^{b_f}$).  This more restrictive ansatz, as compared to the 
standard approach, implies of course less uncertainties \cite{ref6}
concerning the behavior of the parton distributions in the small--$x$
region at $Q >\mu$ which is entirely due to QCD dynamics at 
$x$ \raisebox{-0.1cm}{$\stackrel{<}{\sim}$} $10^{-2}$.  The valence--like
input distributions at $Q_0\equiv \mu < 1$ are parametrized according
to \cite{Gluck:2007sq,ref8}
\begin{eqnarray}
xq_v(x,Q_0^2)&\!\!\!\!\! = &\!\!\!\!\! N_{q_v}x^{a_{q_v}}(1-x)^{b_{q_v}}
      (1+c_{q_v}\sqrt{x} \nonumber \\
&& + d_{q_v}x +e_{q_v}x^{1.5}),\nonumber 
\end{eqnarray}
\begin{equation} 
xw(x,Q_0^2) = N_w x^{a_w}(1-x)^{b_w}(1+c_w\sqrt{x}+d_w x)
\end{equation}
for the valence $q_v=u_v,\, d_v$ and sea $w=\bar{q},\, g$ densities,
and a vanishing strange sea at $Q^2=Q_0^2$, $s(x,Q_0^2)=\bar{s}(x,Q_0^2)=0$.
All further theoretical details relevant for analyzing $F_2$ at NLO and
NNLO in the $\overline{\rm MS}$ factorization scheme have been presented
in \cite{ref8}.  The heavy flavor (dominantly charm) contribution to $F_2$
is taken as given by fixed--order NLO perturbation theory \cite{ref9,ref10}
using $m_c = 1.3$ GeV and $m_b =4.2$ GeV as implied by optimal fits
\cite{ref6} to recent deep inelastic $c$-- and $b$--production HERA
data.  Since a NNLO calculation of heavy quark production is not yet
available, we have again used the same NLO ${\cal{O}}(\alpha_s^2)$
result.  
Finally, we have used for our
fit--analyses the same deep inelastic HERA--H1, BCDMS and NMC data,
with the appropriate cuts for $F_2^{p,n}$ as in \cite{ref8} which
amounts to a total of 740 data points. The required overall 
normalization factors of the data turned out to be 0.98 for H1
and BCDMS, and 1.0 for NMC. We use here again solely deep
inelastic scattering data since we are mainly interested in the
small--$x$ behavior of structure functions.  The resulting parameters
of the NLO and NNLO fits are summarized in Table 1.  The dynamical
gluon and sea distributions, evolved to some specific values of
$Q^2 > Q_0^2$, are at the NLO level very similar to the ones in 
\cite{ref6} which were obtained from a global analysis including
Tevatron Drell--Yan dimuon production and high--E$_T$ inclusive jet
data as well.  Furthermore, the dynamically generated gluon is steeper
as $x\to 0$ than the gluon distributions obtained from conventional
 `standard' fits \cite{ref6,ref8}(based on some arbitrarily chosen
input scale $Q_0^2 > 1$ GeV$^2$, i.e.\ $Q_0^2\simeq 2$ GeV$^2$). On
the other hand, the dynamical sea distribution has a rather similar
small--$x$ dependence as the   `standard' ones \cite{ref6,ref8};
this is caused by the fact that
the valence--like sea input in (2) vanishes very slowly as $x\to 0$
(corresponding to a small value of $a_{\bar{q}}$, $a_{\bar{q}}\simeq
0.05$ according to Table 1) and thus is similarly increasing with 
decreasing $x$ down to $x\simeq 0.01$ as the sea input obtained by
a  `standard' fit.  Similar remarks hold when comparing dynamical and
standard distributions at NNLO.  At NNLO the gluon distribution $xg$
is flatter as $x$ decreases and, in general, falls below the NLO one
in the small--$x$ region, typically by 20 -- 30\% at $x\simeq 10^{-5}$
and 
$Q^2$ \raisebox{-0.1cm}{$\stackrel{<}{\sim}$} $10$ GeV$^2$,
whereas the NNLO sea distribution $x\bar{q}$ is about 10 -- 20\%
larger (steeper) than the NLO one.

\begin{figure}[htb]
\vspace{-2.5cm}
\hspace{-0.6cm}
\epsfig{figure= 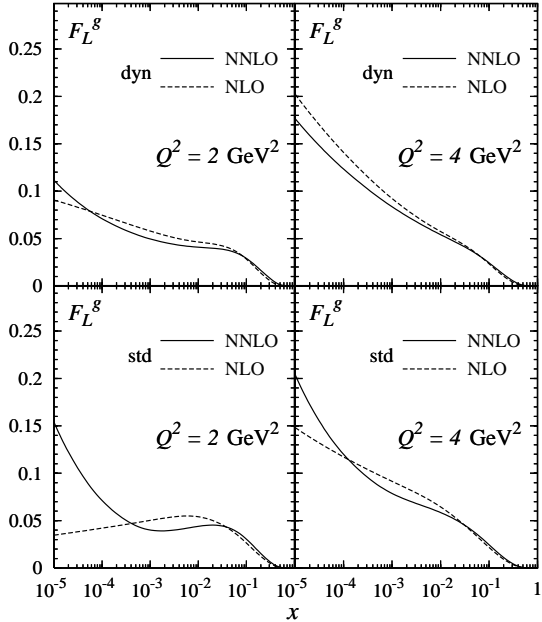, width = 8.cm}
\vspace{-1.5cm}
\caption{The gluonic contribution $F_L^g$ to $F_L$
in (\ref{eq:fl}) with $F_L^g=\frac{2}{9}x\, C_{L,g}\otimes g$ in the 
dynamical (dyn) and
 standard (std) parton
approach at NNLO and NLO for two representative low values of $Q^2$. 
The standard parton distributions utilized in the lower panel are 
taken from \cite{ref8}.}
\vspace{-0.4cm}
\label{fig:flg}
\end{figure}
\begin{figure}[htb]
\vspace{-1.6cm}
\hspace{-0.6cm}
\epsfig{figure= 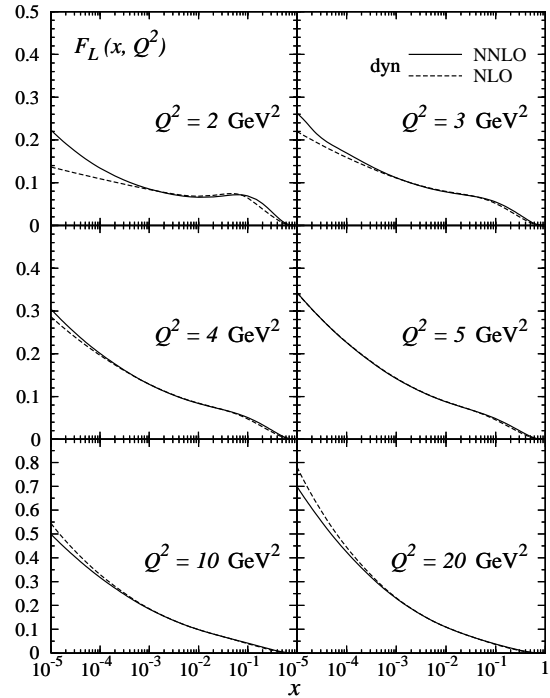, width = 8.cm}
\vspace{-1.5cm}
\caption{Dynamical parton model NNLO and NLO predictions for $F_L(x,Q^2)$.}
\vspace{-0.4cm}
\label{fig:fl_dyn}
\end{figure}
\begin{figure}[htb]
\vspace{-1.6cm}
\hspace{-0.6cm}
\epsfig{figure= 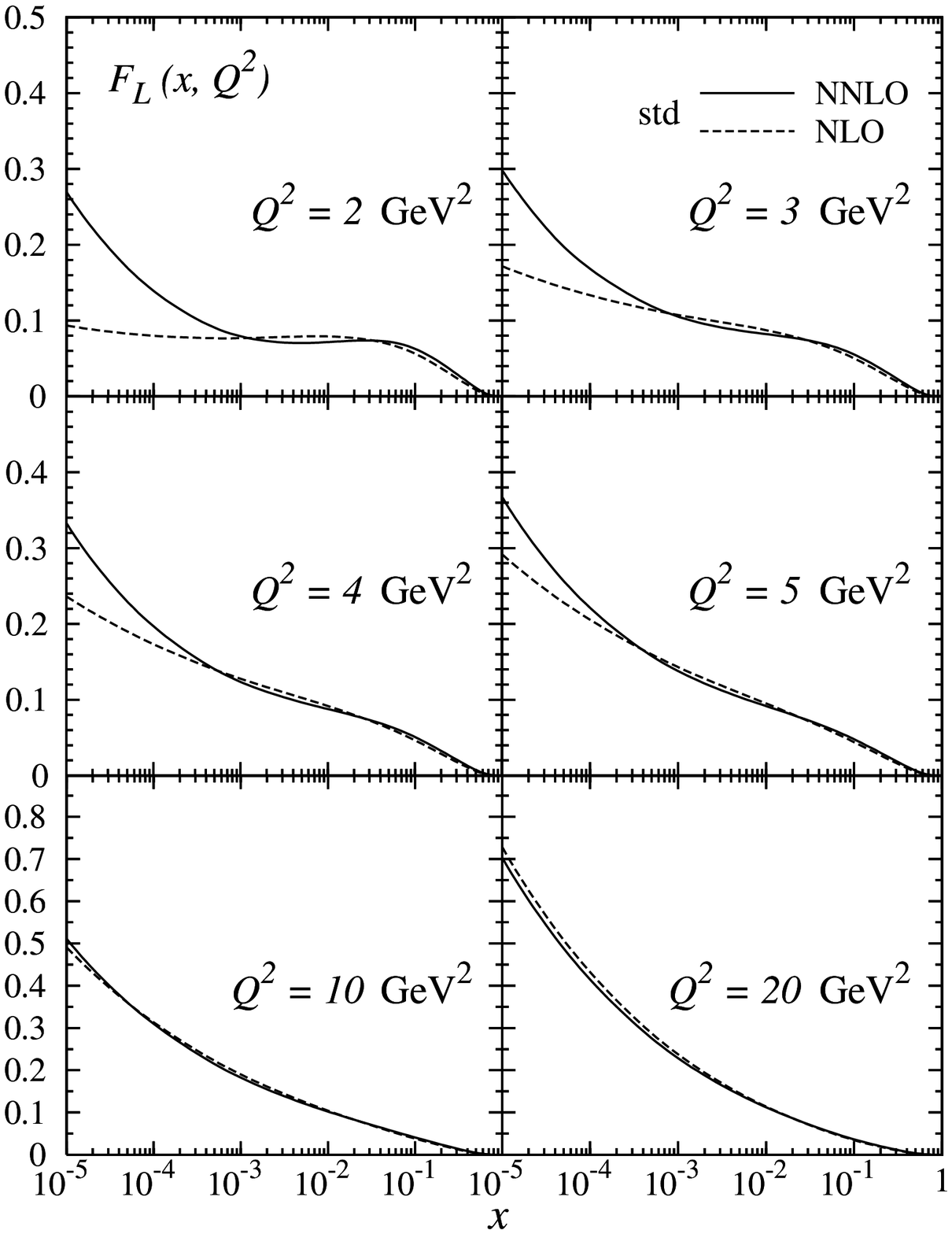, width = 8.cm}
\vspace{-1.5cm}
\caption{As in Fig.~\ref{fig:fl_dyn} but for the common standard parton 
distributions as taken from \cite{ref8}.}
\vspace{-0.4cm}
\label{fig:fl_std}
\end{figure}

Now we turn to the perturbative predictions for $F_L(x,Q^2)$ which
can be written as 
\begin{eqnarray}
x^{-1}F_L & \!\!\!\!= & \!\!\!\!C_{L,ns} \otimes q_{ns} +\frac{2}{9}\,
\left( C_{L,q}\otimes q_s +C_{L,g}\otimes g\right)\nonumber  \\
&& +x^{-1}F_L^c
\label{eq:fl}
\end{eqnarray}
where $\otimes$ in the $n_f=3$ light quark flavor sector denotes the
common convolution, $q_{ns}$ stands for the usual flavor non--singlet
combination and $q_s=\sum_{q=u,d,s}(q+\bar{q})$ is the corresponding
flavor--singlet quark distribution.  Again we use the NLO expression
\cite{ref9,ref10} for $F_L^c$ also in NNLO due to our ignorance of
the ${\cal{O}}(\alpha_s^3)$ NNLO heavy quark corrections.  The
perturbative expansion of the coefficient functions can be written as
\begin{equation}
C_{L,i}(\alpha_s,x) = \sum_{n=1}\, 
  \left( \frac{\alpha_s(Q^2)}{4\pi}\right)^n \, c_{L,i}^{(n)}(x)\, .
\end{equation}
In LO, $c_{L,ns}^{(1)} =\frac{16}{3}x$, $c_{L,ps}^{(1)}=0$,
$c_{L,g}^{(1)}=24x(1-x)$ and the singlet--quark coefficient function
is decomposed into the non--singlet and a `pure singlet' contribution,
$c_{L,q}^{(n)} = c_{L,ns}^{(n)} + c_{L,ps}^{(n)}$.  Sufficiently
accurate simplified expressions for the NLO and NNLO coefficient
functions $c_{L,i}^{(2)}$ and $c_{L,i}^{(3)}$, respectively, have
been given in \cite{ref1}.  It has been futhermore noted in \cite{ref1}
that especially for $C_{L,g}$ both the NLO and NNLO contributions are
rather large over almost the entire $x$--range.  Most striking,
however, is the behavior of both $C_{L,q}$ and $C_{L,g}$ at very
small values of $x$:  the vanishingly small LO parts ($xc_{L,i}^{(1)}
\sim x^2$) are negligible as compared to the (negative) constant
NLO 2--loop terms, which in turn are completely overwhelmed by the
positive NNLO 3-loop singular corrections $xc_{L,i}^{(3)}\sim -\ln x$.
This latter singular contribution might be indicative for the 
perturbative instability at NNLO \cite{ref1}, as discussed at the 
beginning, but it should be kept in mind that a small--$x$ information
alone is {\em{insufficient}} for reliable estimates of the 
convolutions occurring in (\ref{eq:fl}) when evaluating physical observables.

The gluonic contribution $F_L^g$ to $F_L$ in (\ref{eq:fl}) is shown in 
Fig.~\ref{fig:flg} at two characteristic low values of $Q^2$. 
Although the perturbative instability of the subdominant quark contribution 
$F_L^q$ as obtained in a  `standard' fit does not improve for the
dynamical (sea) quark distributions \cite{Gluck:2007sq}, it is evident 
from Fig.~\ref{fig:flg} that the instability disappears 
almost entirely for the dominant dynamical gluon contribution already
at $Q^2\simeq 2$ GeV$^2$.  
This implies that the 
dynamical predictions for the total $F_L(x,Q^2)$ become perturbatively
stable already at the relevant low values of 
$Q^2$ \raisebox{-0.1cm}{$\stackrel{>}{\sim}$} ${\cal{O}}(2-3$ GeV$^2)$
as shown in Fig.~\ref{fig:fl_dyn}, 
in contrast to the  `standard' results in
Fig.~\ref{fig:fl_std}.  In the latter case the stability has not been fully 
reached even at $Q^2 = 5$ GeV$^2$ where the NNLO result at $x=10^{-5}$
is more than 20\% larger than the NLO one.  A similar discrepancy 
prevails for the dynamical predictions in Fig.~3 at $Q^2=2$ GeV$^2$.
This is, however, not too surprising since $Q^2=2$ GeV$^2$ represents
somehow a borderline value for the leading twist--2 contribution to
become dominant at small $x$ values.  This is further corroborated
by the observation that the dynamical NLO twist--2 fit slightly
undershoots the HERA data for $F_2$ at $Q^2 \simeq 2$ GeV$^2$ in the
small--$x$ region (cf.~Fig.~1 of \cite{ref6}).  The NLO/NNLO
instabilities implied by the standard fit results obtained in 
\cite{ref2,ref3} at 
$Q^2$ \raisebox{-0.1cm}{$\stackrel{<}{\sim}$} 5 GeV$^2$ are even
more violent than the ones shown in Fig.~\ref{fig:fl_std}.  
This is mainly due to
the negative longitudinal cross section (negative $F_L(x,Q^2)$) 
encountered in \cite{ref2,ref3}. The perturbative stability in any
scenario becomes in general better the larger $Q^2$, typically beyond
5 GeV$^2$ \cite{ref1,ref2,ref3}, as shown in Figs.~\ref{fig:fl_dyn} and 
\ref{fig:fl_std}.  This is
due to the fact that the $Q^2$--evolutions eventually force any
parton distribution to become sufficiently steep in $x$.

To summarize, we have shown that the extreme perturbative NNLO/NLO
instability of the longitudinal structure function $F_L$ at low $Q^2$,
noted in [2--4], is an artefact of the commonly utilized  `standard'
gluon distributions rather than an indication of a genuine problem
of perturbative QCD.  In fact we have demonstrated that these
extreme instabilities are reduced considerably already at $Q^2 = 2-3$
GeV$^2$ when utilizing the appropriate, dynamically generated, 
parton distributions at NLO and NNLO. These latter parton distributions
have been obtained from a NLO and NNLO analysis of $F_2^{p,n}$ data,
employing the concepts of the dynamical parton model. 
It is gratifying to notice,
once again, the advantage of the dynamical parton model approach to
perturbative QCD.

\section*{Acknowledgments}

This research is part of the 
   research program of the ``Stichting voor Fundamenteel Onderzoek der 
   Materie (FOM)'', which is financially supported by the ``Nederlandse 
   Organisatie voor Wetenschappelijk Onderzoek (NWO)''.

\end{document}